\documentclass[conference]{IEEEtran}
\IEEEoverridecommandlockouts
\usepackage{cite}
\usepackage{amsmath,amssymb,amsfonts}
\usepackage{algorithmic}
\usepackage{graphicx}
\usepackage{textcomp}
\usepackage{xcolor}
\def\BibTeX{{\rm B\kern-.05em{\sc i\kern-.025em b}\kern-.08em
    T\kern-.1667em\lower.7ex\hbox{E}\kern-.125emX}}
\usepackage[hidelinks]{hyperref}
\usepackage{myacronyms}
\usepackage{mytodonotes}
\usepackage{cleveref}
\begin{document}

\title{
    Towards adaptive trajectories for mixed autonomous and human-operated ships
}

\author{\IEEEauthorblockN{Danilo Pianini}
\IEEEauthorblockA{\textit{Department of Computer Science and Engineering} \\
\textit{Alma Mater Studiorum---Università di Bologna}\\
Cesena, Italy \\
0000-0002-8392-5409}
\and
\IEEEauthorblockN{Sven Tomforde}
\IEEEauthorblockA{\textit{Intelligent Systems Group} \\
\textit{Kiel University}\\
Kiel, Germany \\
st@informatik.uni-kiel.de}
}

\maketitle

\begin{abstract}
    We are witnessing the rise of autonomous cars, which will likely revolutionize the way we travel.
    Arguably, the maritime domain lags behind,
    as ships operate on many more degrees of freedom
    (thus, a much larger search space):
    there is less physical infrastructure,
    and rules are less consistent and constraining than what is found on roads.
    The problem is further complicated by the inevitable co-existence of autonomous and human-operated ships:
    the latter may take unpredictable decisions,
    which require adjustments on the autonomous ones.
    Finally,
    the problem is inherently decentralised,
    there is no central authority,
    and communication means can be very diverse in terms of communication distance and performance,
    mandating special care on which information is shared and how.
    In this work,
    we elaborate on the challenges of trajectory prediction and adaptation for mixed autonomous and human-operated ships,
    and we propose initial ideas on potential approaches to address them.
\end{abstract}



\section{Introduction}
\label{sec:intro}

In response to challenges such as staff shortages,
new types of ships and the continuous technical advancement,
considerable effort has recently been made to establish autonomous ships \cite{munim2019autonomous}.
Corresponding technical solutions can currently be observed mainly on the open sea (e.g., as an autopilot)
-- however, harbour and coastal scenarios with correspondingly complex traffic and environmental situations
are also increasingly being focused upon.

A central problem for autonomous shipping is navigation --
which is made up of the sub-problems of general trajectory planning and short-term manoeuvres (e.g., to avoid collisions) \cite{zhang2021collision}.
Corresponding navigation decisions take into account static conditions,
which are available in nautical maps, for example, as well as dynamic influences such as objects or other ships.
Further influences such as mixed autonomous/human operated boats require constant adaptation.
The basis for this is a corresponding perception and classification of the conditions.

As ships also have different communication options,
we propose improving the navigation planning and adaptation problem through a decentralised and self-organised scheme to share local knowledge.
The available communication means change depending on utilisation, position, and disturbances;
for instance, and the environmental conditions change due to shipping behaviour as well as human influences and disturbances such as accidents, damaged ships, flotsam or abnormal traffic including large navy cruisers.
%
We define an open system in which participants can opportunistically use others' local information
to optimise their navigation decisions.
In response to limitations in bandwidth, communication distance or utilisation,
very small amounts of data can be communicated.
We present two basic approaches for a self-organised solution based on machine learning and field-based approaches.

Conceptually, the problem poses a challenge for \ac{SISSY}~\cite{BellmanGLT19,BellmanBDEGLLNP21} approaches,
as each ship autonomously decides on its integration status.
In particular,
it determines independently and depending on the context
which information should and can be shared with its neighbours and which communication basis is used.

The remainder of this article is organised as follows:
Section~\ref{sec:sota} briefly summarises current developments in autonomous ships.
Section~\ref{sec:kiel} introduces the test environment and test vehicle in the Kiel Fjord, Germany,
as a reference scenario.
Based on this,
Section~\ref{sec:proposal} derives two conceptual ways of tackling the problem.
Finally, Section~\ref{sec:conclusion} summarises the article and mentions future work.



\section{Current state in autonomous ships}
\label{sec:sota}

Similar to autonomous cars,
seven levels of autonomy can be defined for ships --
ranging from purely manual operation
to fully autonomous unsupervised solutions,
passing through intermediate steps such as
on-ship decision support,
human-in-the-loop,
and supervised autonomous operations.
For open sea scenarios with very low traffic,
automatic pilots are already in place, reflecting autonomy level 6 out of 7.
However, for environments with moderate or intense traffic,
such as harbours or coastal areas,
no commercial solution exists so far.

Recently,
a variety of projects emerged in academia and industry aiming at the development of technology for autonomous ships,
with the end goal of achieving complete autonomy.
One prominent scenario for this research
(which we will discuss in detail in \Cref{sec:kiel})
is the Kiel Fjord in Germany,
where the main focus is on autonomous ferries.
However, we reckon that a larger market is expected for cargo ships.
For instance,
the Norwegian company Kongsberg Maritime completed one of the first major research projects in this field:
\ac{MUNIN}~\cite{munin}.
Finnish Finferries in cooperation with Rolls-Royce have developed various autonomous vessel prototypes
within the scope of two projects:
\ac{AAWA}~\cite{aawa}
and \ac{SVAN}~\cite{svan}.
Moreover,
there are currently several active projects such as \ac{MAS}~\cite{mas} from IBM
and \ac{AEGIS}~\cite{aegis} by companies and universities that aim to develop solutions
not only for ships but also for logistics and port infrastructure.

One of the most widely used ship autonomy classifications is presented by \ac{MASS}~\cite{mass}
published by the United Kingdom Maritime Autonomous Systems and used by the \ac{IMO}.
According to such classification,
all ship prototypes in the aforementioned projects are not classified as fully autonomous,
because during experiments they were continuously observed by the human operators
with the ability to interrupt the processes at any time and overrule commands,
or even take full control.

Adaptive communication for (semi-)autonomous ships has so far mainly been investigated in coastal areas
with corresponding infrastructure as a basis for monitoring by a \ac{RCC}~\cite{SmirnovT2024}.
Attempts are already being made to make opportunistic use of different available communication mechanisms
(i.e., 4G/5G, Wifi).
Currently, for example, an adaptation of the transmitted sensor information -- here focusing on camera streams --
has been presented using deep reinforcement learning~\cite{SmirnovT23}.
The adaptation takes into account dynamic channel properties and capacity utilisation.
Camera data represents the greatest challenge in comparison to lidar and radar data,
as the data volumes are very large.
A context-related selection of prioritised data has so far only been conceptualised, but not evaluated.
As an alternative direction, adaptive communication has been considered for federated processing of individual ship data for maintenance purposes; consequently, establishing adaptive Internet-of-Ships mechanisms \cite{zhang2021adaptive}. However, this does not include bandwidth usage or navigation purposes.
A scheme for self-organised adaptive communication in open, collective systems, as considered in this paper,
has also not yet been investigated.


\section{Reference future scenario}
\label{sec:kiel}

For illustration purposes,
we use the current developments towards an autonomous ferry in the Kiel Fjord, Germany, as the underlying scenario.
Kiel is a city in northern Germany which is highly characterised by its large harbour and the maritime industry.
Geographically, it is divided into an eastern and a western shore of the fjord.
Hence, public transport mostly relies on busses and ferries.
The fjord itself is also the eastern end of the Kiel Channel -- one of the busiest waterways in Europe.

The fjord is home to a variety of ships -- from the smallest sailing and rowing boats to cruise liners and container ships.
As the largest German naval harbour, fleet movements are also possible.
The water area is divided into different areas:
sailing area,
mooring area,
marine area or fairway for large ships -- all characterised by differing rules and navigability for autonomous ships.
As part of the city, the inner area is very well covered by mobile phone networks up to 5G,
piers are equipped with Wi-Fi and some of the ships have additional uplink options.

\begin{figure}[ht]
\center
\includegraphics[width=0.95\columnwidth]{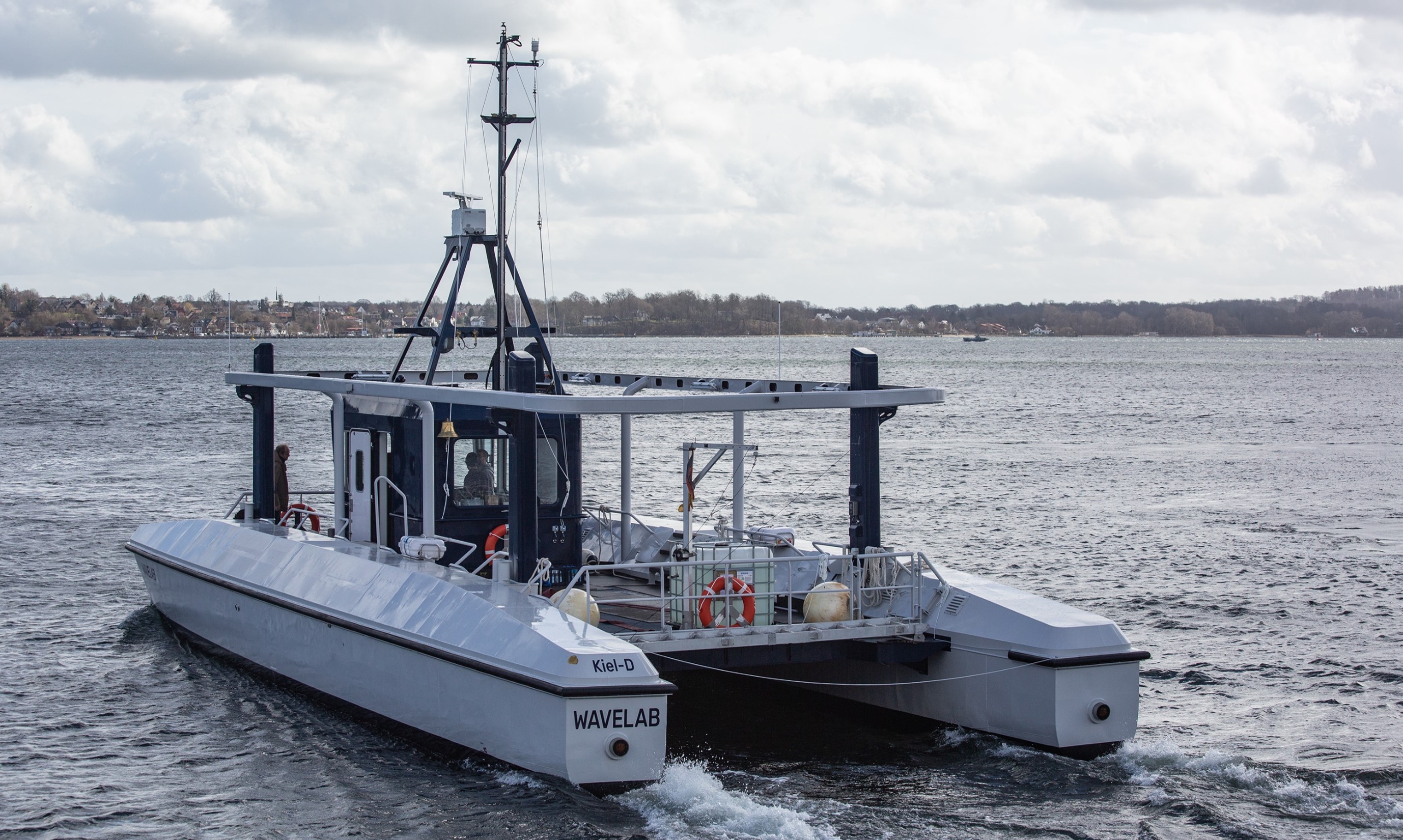}
\caption{The ``MS Wavelab'' -- a test environment for autonomous ferries in Kiel (June 2023).}
\label{fig:wavelab}
\end{figure}

Current developments and research activities are based on the experimental vessel
``MS Wavelab'' (see \Cref{fig:wavelab})~\cite{SchwingerAT23},
equipped with sensors such as several 4k cameras, lidar, radar,
distance sensors and an AIS system.
As autonomous operation is not yet legally permitted, a captain is on board.
In addition, behaviour is monitored from an \ac{RCC}.
For this purpose, as outlined at the beginning,
perception information is already communicated to the RCC in high resolution, depending on the available communication links.

In the following, we will use this scenario as a basis.
Future autonomous ferries -- based on the current MS Wavelab --
will establish public transport on the fjord in regular operation.
They will have to continuously plan trajectories to the landing points according to the timetable
and adapt their short-term navigation behaviour based on the environmental conditions determined by their sensors.
Since this only enables very local situation modelling characterised by occlusions (e.g., by a passing container ship),
we assume that other autonomous ships also continuously communicate their perception as a broadcast --
and that additional information is available for navigation trajectory planning at various levels of detail.

\subsection{Challenges}\label{subsec:challenges}

Navigation of autonomous ferries in the Kiel Fjord poses several challenges.
Although these are similar in other places,
solutions will necessarily need to be adapted,
as navigation rules differ depending on the area.

\paragraph{Heterogeneity}
ships come in extremely different sizes, tonnage, and, consequently, manoeuvrability.
For instance, a large cruise liner will require hundreds of meters to change its course,
and even kilometres to stop:
adaptation must consider the characterising qualities of each ship
to provide viable solutions.

\paragraph{Mixed autonomic and human operation}
many (and, for a long transient period, most) ships in the fjord will be operated by humans,
whose behaviour can be unpredictable.
Moreover, human operators can always override the autonomous system,
thus, even interacting with another autonomous ship
requires the ability to adapt.

\paragraph{Partial information}
different ships are equipped with different sensors,
providing different levels of information.
Additionally,
some ships may be off the grid,
and provide no digital information at all
to other ships.
In this case, those who can perceive them
(through cameras, for instance),
must inform other ships about their presence,
and provide as much information as possible.

\paragraph{Communication challenges}

\begin{figure}
    \centering
    \includegraphics[width=1\columnwidth]{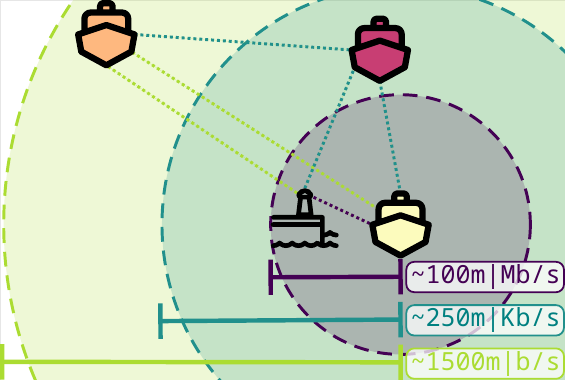}
    \caption{
        Pictorial representation of the communication technologies aboard a ship.
        Different colours indicate a different communication technology:
        short- (purple), mid- (green), and long-range (lime).
        Dashed lines with coloured shades represent communication ranges,
        dotted lines represent communication links.
        Ships may use multiple communication technologies at the same time
        to communicate with the land station or other ships.
    }
    \label{fig:communication-challenges}
\end{figure}

ships are equipped with diverse communication technologies,
implementing different communication ranges and bandwidths.
At short range (about 100m), ships can communicate with each other and with the mainland using Wi-Fi,
while more distant communication (about 250m) is possible using 4G/5G networks, which, however, have a lower data rate.
At very long distances (1km and above), ships may communicate using \ac{LoRa} radio communication;
however, this technology has data rates in the order of tens to a few hundreds bits per second.
\Cref{fig:communication-challenges} illustrates the communication technologies aboard a ship.
The consequences of these heterogeneous communication technologies are twofold:
\begin{enumerate}
    \item data to be exchanged must be dynamically adapted to the communication technology, and
    \item the most appropriate communication channel should be opportunistically
    selected depending on the message recipient.
\end{enumerate}

\paragraph{Safety and security}

the context in which ships operate is highly regulated,
and safety is a primary concern,
as collisions can have catastrophic consequences.
Thus,
validation and verification of any proposed solution are paramount,
and these aspects include both \emph{functional}
(i.e., ensuring that the software behaves as expected)
and \emph{non-functional} verification
(i.e., ensuring that the system is robust to adversarial operations).



\section{Promising research directions}
\label{sec:proposal}

As introduced in \Cref{subsec:challenges},
one of the most prominent challenges to be addressed is the continual modification of the available communication means
and their performance.
In general, direct communication with an \ac{RCC} cannot be taken for granted
(thus, decentralised approaches are required),
yet, when available, it should be leveraged to improve coordination
(consequently, we would like to able to mix complete and partial decentralisation opportunistically).

We believe that this kind of system is particularly ill-suited for traditional approaches
in which the interaction is explicitly designed my means of communication protocols,
as they themselves would be subject to runtime adaptation,
complicating the design considerably.
Instead,
in this paper,
we propose the idea that achieving collective adaptation in these challenging scenarios could be pursued
using macroprogramming~\cite{DBLP:journals/csur/Casadei23} techniques,
and, in particular, with aggregate computing~\cite{BealIEEEComputer2015},
a member of the family of field-based coordination approaches.
In short,
the idea is to design and programme the system as a whole,
programming the ensemble of ships as a single entity,
and letting the compiler/interpreter figure out how to build the interaction and where to distribute the computation.
This way,
we believe it will be possible to partition into separate sub-problems two fundamentally orthogonal aspects of the problem:
\begin{enumerate}
    \item what is the \emph{collective behaviour} we desire ships to achieve, and
    \item how we can opportunistically leverage the best available means to achieve it.
\end{enumerate}

One key aspect of this approach is amenable of several verification and validation techniques:
it is built upon a solid formal foundation~\cite{TOCL2019},
there are ways to build programs that are, by construction~\cite{DBLP:journals/tomacs/ViroliABDP18},
self-stabilising~\cite{DBLP:journals/cacm/Dijkstra74}
and
self-adaptive to device distribution~\cite{TAAS2017}.
In addition,
the approach is maturing a reasonably solid development methodology,
in which simulations are performed first at a more coarse-grained level
to assess the functional behaviour of the system,
and then more details are injected to better study
how the system responds to changes, and, in particular,
to study the transients in detail,
as the formal guarantees that can be obtained by construction
are typically concerning the \emph{eventual} behaviour.

\subsection{Field-based approaches to trajectory computation}

Field-based computation has been used in the past
to compute gradient-based adaptive paths
(for instance, for crowd steering\cite{DBLP:journals/fgcs/CasadeiFPRSV19}):
fields respond promptly to changes in the environment,
behaving similarly to distributed control systems.

Paths computed based on gradients have properties that may be particularly useful in this context:
they can easily be configured to avoid obstacles,
and have been demonstrated to be capable of supporting anticipative behaviour~\cite{DBLP:conf/saso/MontagnaPV12}.
Together, these techniques may lead to paths that bend to avoid collisions,
and that can adapt to react in advance to changes in the intentions of other ships.

For these techniques to be useful in this context,
however,
paths alone are not sufficient:
ships move continuously,
hence,
their velocity must be considered to feed the anticipatory engine.
Moreover,
due to the heterogeneity of the ships,
velocity alone cannot provide a sufficiently clear picture:
inertia and, more generally, capabilities of the ships must be considered as well.

\subsection{Adaptive communication through anisotropic fields}

The problem sketched in \Cref{fig:communication-challenges} requires the ability to dynamically adapt
the data exchanged with neighbours based on the available communication technologies:
on a short range, ships can exchange high-resolution data, including camera streams,
but on larger range the available data rate drops exponentially, down to a few bytes per second.

This scenario is generally not found in the current applications of field-based collective computation,
in fact,
it is normally assumed that devices can communicate with those nearby either with no constraints
or with a fixed data rate and loss.
These assumptions,
however,
do not hold anymore in the context of autonomous ships.
A potential solution to this problem is to use ``anisotropic'' fields,
namely fields whose value is not uniform in all directions.

Typical data exchanges in aggregate computing are performed implicitly by a primitive called \emph{neighbourhood},
which provides the values of the field in the vicinity of every point.
Under the hood, this primitive implies a communication step,
in which the local value of a field is shared with neighbours,
and, consequently, at each point in space it is possible to perceive the field value in the surroundings.
A field built in this way is \emph{isotropic}:
regardless of the direction from which the field is observed,
its value is the same---in other words, devices send the same data to all their neighbours.

In recent work,
a different fundamental primitive, called \emph{exchange}~\cite{DBLP:journals/jss/AudritoCDSV24}, has been proposed.
Without delving into the details,
exchange enables sending different data to different neighbours,
and, consequently, to build fields whose value is different in different directions---anisotropic fields.
For the scenario at hand,
this primitive seems to provide a natural way to model fields that dynamically adapt to the data rate of the communication link
of each neighbour.

Adopting this approach
may impact relevantly on the current design of the aggregate programming high-level \acp{API}:
in fact, typically, complex computations are performed by combining high-level,
self-stabilising building blocks~\cite{DBLP:journals/tomacs/ViroliABDP18},
and rarely low-level operators such as neighbourhood and exchange are used directly.
For the approach to be practically viable in the context at hand,
a high-level, self-stabilising \ac{API} must be present,
and whether or not, or to what extent, the current \ac{API} can be extended to support anisotropic fields is an open question.

\subsection{Information dependability through error-aware fields}

An approach that can either complement or substitute the anisotropic fields approach is the use of error-aware fields.
In this case,
the information carried along fields must also include a measure of its reliability,
which gets automatically updated based on the path the information followed to get to its destination.
This way,
error would accumulate,
and data that is considered too unreliable for the current scope of the computation
(either because it took too long to be delivered, or because the error accumulated over a threshold)
could get discarded
or,
if the technique is used in conjunction with the anisotropic fields,
not even transmitted in the first place.

The introduction of error-aware fields is another change with the potential to impact significantly
on the way aggregate computing is performed.
Research is required to understand if and how it could be captured at the type-system level,
and if such change could be limited to the introduction of new types or will impact the set of available operators as well.

\subsection{Runtime reconfiguration via pulverisation}

One of the central challenges in autonomous ship control is the extreme heterogeneity of the involved devices.
As introduced in \Cref{subsec:challenges},
ships are extremely diverse, and diversity can also involve the digital equipment on board.
For instance,
while the prototype of the MS Wavelab is equipped with a variety of sensors and remarkable computational power
(with additional computational power available on the land station),
the same cannot be said for all ships in the fjord.
However, when designing control software for such a complex system,
taking into account all the unavoidable heterogeneity is a daunting task.
Ideally, the designer would like to specify their control software once,
and have it automatically adapt to the available computational resources,
depending on the current circumstances.

Pulverisation is a recent technique that promises to achieve this goal~\cite{DBLP:journals/fi/CasadeiPPVW20,DBLP:journals/iotj/CasadeiFPPSV22}.
The core of the idea is to separate the computation into micro-activities
whose deployment can be decided independently of the functional definition of the system,
and, possibly, reconfigured at runtime.
This way, the designer can reason about the system as a whole,
considering all ships as if they were equipped with all the machinery required to perform the computation,
and let the system decide how to distribute the computation based on the available resources.
Then, after pulverisation has been applied,
portions of the behaviour of a ship can be hosted on different devices,
such as nearby
(better equipped)
ships, land stations, or even the cloud, if the ship can utilise an Internet connection.

Aggregate computing is one of the macro-programming approaches that is ``naturally pulverisable'',
meaning that pulverisation can be applied with no special needs to adapt the programming model.
In the literature,
pulverisation has been applied to aggregate systems by splitting them into five subcomponents:
\begin{enumerate}
    \item sensors;
    \item actuators;
    \item behaviour;
    \item communication;
    and
    \item state.
\end{enumerate}
Of them,
the former two have strict bounds regarding the devices they can be hosted on,
at least in the context of autonomous ships,
as they would provide sensor readings and actuation commands,
which are inherently tied to the physical ship.
Still,
the other three components can be pulverised and distributed to different devices:
for instance,
the behaviour of a ship can be hosted on a nearby ship,
for instance because the trajectory computation requires more computational power than the local ship can provide;
or because the other ship is close to a land station
(with which it can communicate using a high-speed connection),
but can only use a low-speed connection to communicate with the local ship.
In the latter case,
since only raw sensor readings (such as position and heading)
need to be exchanged between sensor and behaviour,
and all application-level communication is performed on a nearby ship
(thus, possibly, in-memory),
the packet size may be minimal.

Applying this technique to the problem at hand,
especially in conjunction with other promising research directions
we introduced before,
is likely to promote the development of a refined pulverisation technique
for aggregate computing.
Consider, for instance,
the case of a ship equipped with several high resolution cameras and lidars,
but whose connection to the nearest ship or land station is very slow:
applying the aforementioned technique and moving away processing to share solely raw data with the nearby processing device
would be counterproductive,
as in this case the raw data produced by sensors is larger than any coordination data that could be exchanged
as a result of the computation.
In this case,
we probably want to pulverise our system differently,
with a ``data process'' component and a ``coordination behaviour'' component,
of which the first is in charge of producing a compact summary of the information
from raw data
through a preliminary processing step,
while the latter computes the coordination acts and the final trajectory based on the summary data.
This way,
if we have good communication with a better-equipped device we could move the data processing there
and send raw data,
while, in case the connection is too slow,
we could keep the data processing on board and send the summary data instead.

As per the other research directions,
novel techniques open to innovative solutions for the challenges of the problem at hand,
and, at the same time,
their application to concrete examples provides ideas to refine and improve the techniques themselves.

\subsection{Learning-based approaches}

Recent advances in field based computation have fostered the idea of using learning-based approaches to compute fields~\cite{DBLP:conf/icdcs/AguzziCV22}.
In these cases, multiple simulation rounds are performed to learn the collective behaviour
(as the experiences need to be collected in a centralised fashion)
which is then executed in a fully distributed way.
Crucially,
as far as a suitable reward function can be defined,
the approach is general enough to be applied to multiple problems,
including trajectory computation, ships behaviour prediction, and selection of the most crucial data to be exchanged.

These approaches are relatively new, and come with several research challenges that need to be addressed,
especially before they can be applied to the problem at hand.
Among them,
explainability of the collective behaviour is particularly relevant,
as the reasons behind the decisions taken by the system must be clear to the human operators.
A particularly promising approach in this context is \emph{sketching},
in which part of the computation is produced by designers,
who build a template whose holes are filled by the learning algorithm.
This way,
the most critical parts of the computation can remain under human control,
and can be inspected, understood, and certified.


\section{Conclusion}
\label{sec:conclusion}

In this paper,
we have discussed the challenges of trajectory prediction and adaptation for mixed autonomous and human-operated ships
and have proposed initial ideas on potential approaches to address them.
We have introduced the Kiel Fjord as a reference scenario,
and we have discussed the challenges posed by the heterogeneity of the ships,
the mixed autonomy,
the partial information,
and the communication challenges.
We have proposed to address these challenges using field-based computation
and have introduced the idea of anisotropic fields,
error-aware fields,
and pulverisation as potential way to address the challenges of this peculiar scenario.

The proposed approaches are still at a conceptual stage,
and further research is required to understand if and how they can be applied to the problem at hand:
we plan to investigate the feasibility and performance of the proposed approaches
in the near future.


\bibliographystyle{IEEEtran}
\bibliography{bibliography}

\begin{thebibliography}{10}
\providecommand{\url}[1]{#1}
\csname url@samestyle\endcsname
\providecommand{\newblock}{\relax}
\providecommand{\bibinfo}[2]{#2}
\providecommand{\BIBentrySTDinterwordspacing}{\spaceskip=0pt\relax}
\providecommand{\BIBentryALTinterwordstretchfactor}{4}
\providecommand{\BIBentryALTinterwordspacing}{\spaceskip=\fontdimen2\font plus
\BIBentryALTinterwordstretchfactor\fontdimen3\font minus
  \fontdimen4\font\relax}
\providecommand{\BIBforeignlanguage}[2]{{%
\expandafter\ifx\csname l@#1\endcsname\relax
\typeout{** WARNING: IEEEtran.bst: No hyphenation pattern has been}%
\typeout{** loaded for the language `#1'. Using the pattern for}%
\typeout{** the default language instead.}%
\else
\language=\csname l@#1\endcsname
\fi
#2}}
\providecommand{\BIBdecl}{\relax}
\BIBdecl

\bibitem{munim2019autonomous}
Z.~H. Munim, ``Autonomous ships: a review, innovative applications and future
  maritime business models,'' in \emph{Supply Chain Forum: An International
  Journal}, vol.~20, no.~4.\hskip 1em plus 0.5em minus 0.4em\relax Taylor \&
  Francis, 2019, pp. 266--279.

\bibitem{zhang2021collision}
X.~Zhang, C.~Wang, L.~Jiang, L.~An, and R.~Yang, ``Collision-avoidance
  navigation systems for maritime autonomous surface ships: A state of the art
  survey,'' \emph{Ocean Engineering}, vol. 235, p. 109380, 2021.

\bibitem{BellmanGLT19}
K.~L. Bellman, C.~Gruhl, C.~Landauer, and S.~Tomforde, ``Self-improving system
  integration - on a definition and characteristics of the challenge,'' in
  \emph{{IEEE} 4th International Workshops on Foundations and Applications of
  Self* Systems, FAS*W@SASO/ICCAC 2019, Umea, Sweden, June 16-20, 2019}.\hskip
  1em plus 0.5em minus 0.4em\relax {IEEE}, 2019, pp. 1--3.

\bibitem{BellmanBDEGLLNP21}
K.~L. Bellman, J.~Botev, A.~Diaconescu, L.~Esterle, C.~Gruhl, C.~Landauer,
  P.~R. Lewis, P.~R. Nelson, E.~Pournaras, A.~Stein, and S.~Tomforde,
  ``Self-improving system integration: Mastering continuous change,''
  \emph{Future Gener. Comput. Syst.}, vol. 117, pp. 29--46, 2021.

\bibitem{munin}
``{Maritime Unmanned Navigation through Intelligence in Networks (MUNIN}),''
  \url{https://bit.ly/3XKAlxF}, accessed on 18.04.2024.

\bibitem{aawa}
``{Advanced Autonomous Waterborne Applications (AAWA}),''
  \url{https://bit.ly/3JP6deQ}, accessed on 18.04.2024.

\bibitem{svan}
``{Safer Vessel with Autonomous Navigation (SVAN}),''
  \url{https://breakingwaves.fi/wp-content/uploads/2019/06/SVAN-presentation.pdf},
  accessed on 18.04.2024.

\bibitem{mas}
``{Mayflower Autonomous Ship (MAS}),'' \url{https://mas400.com/}, accessed on
  18.04.2024.

\bibitem{aegis}
``{Advanced, Efficient and Green Intermodal Systems (AEGIS}),''
  \url{https://aegis.autonomous-ship.org/}, accessed on 18.04.2024.

\bibitem{mass}
``{Maritime Autonomous Ship Systems (MASS) UK Industry Conduct Principles and
  Code of Practice, version 4, November 2020},'' \url{https://bit.ly/3O6q1gj},
  accessed on 18.04.2024.

\bibitem{SmirnovT2024}
\BIBentryALTinterwordspacing
N.~Smirnov and S.~Tomforde, ``Real-time rate control of webrtc video streams in
  5g networks: Improving quality of experience with deep reinforcement
  learning,'' \emph{Journal of Systems Architecture}, vol. 148, p. 103066,
  2024. [Online]. Available:
  \url{https://www.sciencedirect.com/science/article/pii/S1383762124000031}
\BIBentrySTDinterwordspacing

\bibitem{SmirnovT23}
------, ``Real-time data transmission optimization on 5g remote-controlled
  units using deep reinforcement learning,'' in \emph{Architecture of Computing
  Systems - 36th International Conference, {ARCS} 2023, Athens, Greece, June
  13-15, 2023, Proceedings}, ser. Lecture Notes in Computer Science, G.~I.
  Goumas, S.~Tomforde, J.~Brehm, S.~Wildermann, and T.~Pionteck, Eds., vol.
  13949.\hskip 1em plus 0.5em minus 0.4em\relax Springer, 2023, pp. 281--295.

\bibitem{zhang2021adaptive}
Z.~Zhang, C.~Guan, H.~Chen, X.~Yang, W.~Gong, and A.~Yang, ``Adaptive
  privacy-preserving federated learning for fault diagnosis in internet of
  ships,'' \emph{IEEE Internet of Things Journal}, vol.~9, no.~9, pp.
  6844--6854, 2021.

\bibitem{SchwingerAT23}
R.~Schwinger, G.~Al{-}Falouji, and S.~Tomforde, ``Autonomous ship collision
  avoidance trained on observational data,'' in \emph{Architecture of Computing
  Systems - 36th International Conference, {ARCS} 2023, Athens, Greece, June
  13-15, 2023, Proceedings}, ser. Lecture Notes in Computer Science, G.~I.
  Goumas, S.~Tomforde, J.~Brehm, S.~Wildermann, and T.~Pionteck, Eds., vol.
  13949.\hskip 1em plus 0.5em minus 0.4em\relax Springer, 2023, pp. 296--310.

\bibitem{DBLP:journals/csur/Casadei23}
\BIBentryALTinterwordspacing
R.~Casadei, ``Macroprogramming: Concepts, state of the art, and opportunities
  of macroscopic behaviour modelling,'' \emph{{ACM} Comput. Surv.}, vol.~55,
  no. 13s, pp. 275:1--275:37, 2023. [Online]. Available:
  \url{https://doi.org/10.1145/3579353}
\BIBentrySTDinterwordspacing

\bibitem{BealIEEEComputer2015}
\BIBentryALTinterwordspacing
J.~Beal, D.~Pianini, and M.~Viroli, ``Aggregate programming for the internet of
  things,'' \emph{{IEEE} Computer}, vol.~48, no.~9, pp. 22--30, 2015. [Online].
  Available: \url{http://dx.doi.org/10.1109/MC.2015.261}
\BIBentrySTDinterwordspacing

\bibitem{TOCL2019}
\BIBentryALTinterwordspacing
G.~Audrito, M.~Viroli, F.~Damiani, D.~Pianini, and J.~Beal, ``A higher-order
  calculus of computational fields,'' \emph{{ACM} Transactions on Computational
  Logic}, vol.~20, no.~1, pp. 1--55, jan 2019. [Online]. Available:
  \url{https://doi.org/10.1145/3285956}
\BIBentrySTDinterwordspacing

\bibitem{DBLP:journals/tomacs/ViroliABDP18}
\BIBentryALTinterwordspacing
M.~Viroli, G.~Audrito, J.~Beal, F.~Damiani, and D.~Pianini, ``Engineering
  resilient collective adaptive systems by self-stabilisation,'' \emph{{ACM}
  Trans. Model. Comput. Simul.}, vol.~28, no.~2, pp. 16:1--16:28, 2018.
  [Online]. Available: \url{https://doi.org/10.1145/3177774}
\BIBentrySTDinterwordspacing

\bibitem{DBLP:journals/cacm/Dijkstra74}
\BIBentryALTinterwordspacing
E.~W. Dijkstra, ``Self-stabilizing systems in spite of distributed control,''
  \emph{Commun. {ACM}}, vol.~17, no.~11, pp. 643--644, 1974. [Online].
  Available: \url{https://doi.org/10.1145/361179.361202}
\BIBentrySTDinterwordspacing

\bibitem{TAAS2017}
\BIBentryALTinterwordspacing
J.~Beal, M.~Viroli, D.~Pianini, and F.~Damiani, ``Self-adaptation to device
  distribution in the internet of things,'' \emph{ACM Trans. Auton. Adapt.
  Syst.}, vol.~12, no.~3, pp. 12:1--12:29, Sep. 2017. [Online]. Available:
  \url{http://doi.acm.org/10.1145/3105758}
\BIBentrySTDinterwordspacing

\bibitem{DBLP:journals/fgcs/CasadeiFPRSV19}
\BIBentryALTinterwordspacing
R.~Casadei, G.~Fortino, D.~Pianini, W.~Russo, C.~Savaglio, and M.~Viroli,
  ``Modelling and simulation of opportunistic iot services with aggregate
  computing,'' \emph{Future Gener. Comput. Syst.}, vol.~91, pp. 252--262, 2019.
  [Online]. Available: \url{https://doi.org/10.1016/j.future.2018.09.005}
\BIBentrySTDinterwordspacing

\bibitem{DBLP:conf/saso/MontagnaPV12}
\BIBentryALTinterwordspacing
S.~Montagna, D.~Pianini, and M.~Viroli, ``Gradient-based self-organisation
  patterns of anticipative adaptation,'' in \emph{Sixth {IEEE} International
  Conference on Self-Adaptive and Self-Organizing Systems, {SASO} 2012, Lyon,
  France, September 10-14, 2012}.\hskip 1em plus 0.5em minus 0.4em\relax {IEEE}
  Computer Society, 2012, pp. 169--174. [Online]. Available:
  \url{https://doi.org/10.1109/SASO.2012.25}
\BIBentrySTDinterwordspacing

\bibitem{DBLP:journals/jss/AudritoCDSV24}
\BIBentryALTinterwordspacing
G.~Audrito, R.~Casadei, F.~Damiani, G.~Salvaneschi, and M.~Viroli, ``The
  exchange calculus {(XC):} {A} functional programming language design for
  distributed collective systems,'' \emph{J. Syst. Softw.}, vol. 210, p.
  111976, 2024. [Online]. Available:
  \url{https://doi.org/10.1016/j.jss.2024.111976}
\BIBentrySTDinterwordspacing

\bibitem{DBLP:journals/fi/CasadeiPPVW20}
\BIBentryALTinterwordspacing
R.~Casadei, D.~Pianini, A.~Placuzzi, M.~Viroli, and D.~Weyns, ``Pulverization
  in cyber-physical systems: Engineering the self-organizing logic separated
  from deployment,'' \emph{Future Internet}, vol.~12, no.~11, p. 203, 2020.
  [Online]. Available: \url{https://doi.org/10.3390/fi12110203}
\BIBentrySTDinterwordspacing

\bibitem{DBLP:journals/iotj/CasadeiFPPSV22}
\BIBentryALTinterwordspacing
R.~Casadei, G.~Fortino, D.~Pianini, A.~Placuzzi, C.~Savaglio, and M.~Viroli,
  ``A methodology and simulation-based toolchain for estimating deployment
  performance of smart collective services at the edge,'' \emph{{IEEE} Internet
  Things J.}, vol.~9, no.~20, pp. 20\,136--20\,148, 2022. [Online]. Available:
  \url{https://doi.org/10.1109/JIOT.2022.3172470}
\BIBentrySTDinterwordspacing

\bibitem{DBLP:conf/icdcs/AguzziCV22}
\BIBentryALTinterwordspacing
G.~Aguzzi, R.~Casadei, and M.~Viroli, ``Machine learning for aggregate
  computing: a research roadmap,'' in \emph{42nd {IEEE} International
  Conference on Distributed Computing Systems, {ICDCS} Workshops, Bologna,
  Italy, July 10, 2022}.\hskip 1em plus 0.5em minus 0.4em\relax {IEEE}, 2022,
  pp. 119--124. [Online]. Available:
  \url{https://doi.org/10.1109/ICDCSW56584.2022.00032}
\BIBentrySTDinterwordspacing

\end{thebibliography}

\vspace{12pt}

\end{document}